\documentclass[trackchanges, twocolumn]{aastex701}

\usepackage{chemformula}

\begin{document}

\title{\emph{Reflation}: redox-driven atmospheric inflation as tracer of super-Earth geochemistry}

\author[orcid=0009-0009-7228-7809]{Lorenzo Cesario}
\affiliation{Kapteyn Astronomical Institute, University of Groningen, 9747 AD Groningen, The Netherlands}
\email[show]{lorenzo.cesario.research@gmail.com}  

\author[orcid=0000-0002-3286-7683]{Tim Lichtenberg} 
\affiliation{Kapteyn Astronomical Institute, University of Groningen, 9747 AD Groningen, The Netherlands}
\email[show]{tim.lichtenberg@rug.nl}

\author[orcid=0000-0002-7971-7439]{Mara Attia} 
\affiliation{Kapteyn Astronomical Institute, University of Groningen, 9747 AD Groningen, The Netherlands}
\email{m.attia@rug.nl}

\author[orcid=0000-0002-8368-4641]{Harrison Nicholls}
\affiliation{Institute of Astronomy, University of Cambridge, Madingley Road, Cambridge CB3 0HA, UK}
\email{harrison.nicholls@ast.cam.ac.uk}

\author[orcid=0009-0009-7323-6755]{Imre Kisvárdai} 
\affiliation{Kapteyn Astronomical Institute, University of Groningen, 9747 AD Groningen, The Netherlands}
\email{kisvardai@astro.rug.nl}

\author[orcid=0000-0001-6516-4493]{Quentin Changeat}
\affiliation{Kapteyn Astronomical Institute, University of Groningen, 9747 AD Groningen, The Netherlands}
\email{q.changeat@rug.nl}

\begin{abstract}
\noindent We demonstrate that the redox-sensitivity of mantle outgassing can trigger transient episodes of atmospheric re-inflation in highly irradiated and geochemically-reduced super-Earths, a mechanism we term \emph{reflation}. Mantle redox governs the outgassing and speciation of CHONS volatiles, setting the background secondary atmospheric composition during extended photoevaporation at highly irradiated conditions. Using simulations of the coupled atmosphere-interior evolution of irradiated super-Earths, we illustrate that reduced mantles close to the iron-w\"ustite buffer initially produce CO-dominated atmospheres. Hydrodynamic escape continuously removes volatiles while outgassing from the melt replenishes the atmosphere with H$_2$, converted from H$_2$O dissolved in the underlying magma ocean. This leads to a late-stage transition from C- to H-dominated gas that transiently re-inflates super-Earth atmospheres and decreases their bulk densities by up to $\sim$ 60\% between several hundreds of Myr to Gyr after their formation, prior to complete atmospheric erosion by photoevaporation. In contrast, oxidised mantles, closer to Earth-like geochemistry, strongly buffer their atmospheric composition while exposed to hydrodynamic escape, producing monotonic radius deflation. Reflation events are triggered by geochemically-reduced mantles, intermediate escape efficiencies, high irradiation, and initial water inventories $\gtrsim 5$~Earth~oceans. This redox-dependent evolutionary divergence hinges on the sensitive feedback between interior and atmospheric evolution serving as a potential tracer of historical geochemical state. Population-level reflation signatures of close-in super-Earths may thus serve as tracers of interior geochemistry and formation conditions.
\end{abstract}

\section{Introduction}

What determines whether a rocky exoplanet retains an atmosphere, and what that atmosphere is made of? Transit surveys have uncovered a striking bimodality in the radius distribution of close-in planets \citep{Fulton2018AJ, Luque2022Science}: gas-rich sub-Neptunes above $R_p\approx 1.8\,R_\oplus$ and super-Earths below, separated by a `radius valley' commonly attributed to atmospheric escape \citep{Owen2017ApJ, Gupta2019MNRAS}. Yet the fate of the stripped `cores' remains unclear. Some may be truly devolatilised remnants \citep{Rogers2015ApJ}, others may hold onto, or re-generate, secondary atmospheres through degassing from their interiors \citep{Gaillard2021SSRev,Hu2023ApJL}. Distinguishing between these outcomes is now within reach of JWST and upcoming missions \citep{Bean2021JGRP, Greene2023Nature, Kempton2024RMG}, but interpreting the observations requires understanding how interior geochemistry shapes atmosphere climates and observables over Gyr timescales \citep{Wordsworth2022ARAA, Lichtenberg2025TrGeo}. Recent observational studies have found some of the most irradiated super-Earths to be inconsistent with bare-rock surfaces \citep{Zieba2022AA, Hu2024Natur, August2025AA, Monaghan2025AJ, Teske2025ApJL, ParkCoy2026arXiv}. These findings deviate from what one would expect if hydrodynamic escape is the dominant driver of atmospheric evolution of highly irradiated super-Earths. 

A central, yet poorly constrained, control over planetary observables is mantle redox geochemistry, which sets the solubility of volatile species dissolved within silicate melts and comprising outgassed atmospheres \citep{Gaillard2021SSRev, Sossi2023EPSL, Frost2018RMG}. Oxidised mantles tend to outgas heavier CO$_2$- and H$_2$O-dominated atmospheres, while reduced mantles favour lighter CO and H$_2$ compositions \citep{Nicholls2024JGRP, Ortenzi2020SciRep, Gaillard2022EPSL}. Mantle redox state is thus directly correlated with atmospheric mean molecular weight (MMW), scale height, transit depth, and inferred bulk density of observable exoplanets \citep{Dorn2021ApJL, Lichtenberg2021ApJL,Boer2025ApJ,Cherubim2025ApJ,Bower2025ApJ,Nicholls2026NatAs}. On highly irradiated orbits, efficient atmospheric escape processes operate on timescales comparable to interior cooling \citep{Owen2019AREPS, Yoshida2024PEPS}, yet the coupled feedback between redox-controlled outgassing and composition-dependent mass loss is currently lacking clear observational predictions to distinguish between potential evolution pathways \citep{Kempton2024RMG,Lichtenberg2025Science}.

Here we use a coupled interior--atmosphere evolution model to show that the interplay between redox-controlled outgassing and hydrodynamic escape leads to transient, late-stage atmospheric re-inflation enabled by geochemically-driven shifts in atmospheric composition, which we term \emph{reflation}. Rather than monotonically stripping a planet's secondary envelope, removal of volatile elements by hydrodynamic mass-loss enables a compositional transformation which temporarily inflates a planet's observable radius. This only takes place at reducing interior states, but not at oxidized, Earth-like internal geochemistry. Our results thus suggest that the atmosphere presents an evolving record of the deep interior redox geochemistry. We present case studies and map the preferred parameter conditions under which reflation occurs, then discuss the limitations of our employed methods, and the potential implications of the reflation mechanism for observational facilities to test.

\section{Methods}
\label{sec:methods}

\subsection{Coupled interior-atmosphere framework}
\label{sec:methods_proteus}

We simulate the coupled interior-atmosphere evolution of super-Earth planets across a physically-informed parameter space using the modular \textsc{proteus}\footnote{\url{https://proteus-framework.org}} framework \citep{Lichtenberg2021JGRP, Nicholls2024JGRP, Nicholls2024MNRAS}. Starting from a fully-molten state, \textsc{proteus} time-marches planetary evolution while self-consistently facilitating columnar interactions between physical, chemical, and dynamical processes by sharing state variables (e.g. surface temperature, melt fraction, atmospheric composition, radiation fluxes) until atmospheric escape occurs. 
 
These planets are simulated with a conceptual three-domain approach, each of which is handled by one of the `submodules' within the \textsc{proteus} framework.  The \emph{interior} domain, handled  by \textsc{spider} \citep{Bower2018PEPI, Bower2019AA, Bower2022PSJ}, solves the interior structure and energy continuity equations to evolve radial profiles of energy fluxes, mantle temperature, melt fraction, and viscosity over time. The \emph{atmospheric} domain, which is outgassed at the interior-atmosphere surface, is solved using radiative-convective climate calculations by \textsc{agni} \citep[Section~\ref{sec:methods_atmos}]{Nicholls2025JOSS, Nicholls2024MNRAS}. The \emph{stellar} domain resolves the stellar radiation impinging upon the planet, spectroscopically and temporally, using \textsc{mors}  \citep{Johnstone2021AA, Spada2013ApJ}. Section~\ref{sec:methods_escape} details our  energy-limited hydrodynamic atmospheric escape calculations \citep{Lopez2014ApJ, Lehmer2017ApJ, Hunten1987Icarus}. 
 
Atmospheric physics and chemistry largely act on shorter timescales than mantle geodynamics, while atmospheric blanketing remains key for regulating planetary cooling rates \citep{Nicholls2024JGRP, ElkinsTanton2008EPSL, Abe1986JGR}. Here, evolution is time-marched iteratively with the atmosphere and stellar domains. The interior domain is subject to an energy flux boundary condition calculated from the atmosphere, ensuring a full physical feedback loop. \textsc{proteus} is participating in the undergoing benchmark and intercomparison project \textsc{chili} \citep{Lichtenberg2025arXiv}, part of the CUISINES exoplanet model intercomparison framework \citep{Sohl2024}, which provides a controlled baseline for comparing simulations of planetary evolution.

\subsection{Volatile outgassing and climate model}
\label{sec:methods_atmos}

At each time-step of \textsc{proteus}, the partial pressures of the volatile species \ch{H2O}, \ch{H2}, \ch{CH4}, \ch{CO2}, \ch{CO}, \ch{N2}, \ch{NH3}, \ch{SO2}, \ch{H2S}, \ch{S2}, and \ch{O2} at the atmosphere-interior (surface) interface are calculated using the \textsc{calliope} submodule \citep{Bower2019AA, Bower2022PSJ, Sossi2023EPSL, Nicholls2024JGRP, Nicholls2024MNRAS}. Given a surface temperature $T_s$ and the total abundance of CHNS elements ($M_e^\mathrm{tot}$) distributed across the mantle plus atmosphere system, \textsc{calliope} solves for the elemental partitioning between the atmosphere and interior reservoirs subject to mass conservation: $M_e^\mathrm{atm} + M_e^\mathrm{int} = M_e^\mathrm{tot}$. The distribution of oxygen atoms is treated separately and is not conserved across different simulations, because the oxygen is fundamentally coupled to the redox chemistry buffering throughout the mantle \citep{Sossi2020SciAdv,Lichtenberg2021ApJL}. We set the upper-mantle oxygen fugacity $f\ch{O2}$ -- treated as equal to the partial pressure of \ch{O2} -- relative to the temperature-dependent iron--w\"ustite buffer reaction \citep{ONeill2002ChemGeo}, which implicitly resolves the disproportionation of Fe after core segregation. The mass of each CHNOS element in the \textit{atmosphere} reservoir is then obtained by summing over all partial surface pressures: $M_e^\mathrm{atm} = 4 \pi R_\mathrm{int}^2 g\sum_j p_j(\mu_j/\mu)$. Thermochemical reactions between the gases are modelled using temperature-dependent equilibrium coefficients $K_\mathrm{eq}$ for each combination of species \citep{Nicholls2024MNRAS, Chase1986NIST, Schaefer2017ApJ}. Simultaneously, the dissolved mass of each element is obtained by summing over the corresponding dissolved phases; e.g. \ch{OH-} for \ch{H2O} \citep{Nicholls2024JGRP, Bower2022PSJ}. All volatiles partition at thermochemical-solubility equilibrium using empirical solubility laws to relate their partial pressures $p_j$ to dissolved phases' concentrations $x_j$ \citep{Sossi2023EPSL, Gaillard2022EPSL, Dixon1997AmMin, Armstrong2015GCA, Ardia2013GCA, Dasgupta2022GCA}. Outgassed compositions are strongly coupled to the interior geochemistry, because volatile solubilities and chemical equilibria sensitively depend on the mantle redox state -- proxied by $f\ch{O2}$ -- in addition to temperature and pressure \citep{Gaillard2022EPSL, Ortenzi2020SciRep, Schaefer2017ApJ}.

The planet's cooling rate, photospheric radius, and bulk density are obtained by modelling the atmosphere's radiative-convective structure at each time-step \citep{Nicholls2025JOSS, Nicholls2024MNRAS}. Our models assume that the atmosphere attains a steady state on time-scales shorter than the interior geodynamics, so the surface temperatures and partial pressures represent input parameters to the climate calculations, from the preceding partitioning-outgassing solution. We use \textsc{agni} to solve for radiative-convective equilibrium and the total cooling rate of the planet by enforcing energy flux conservation throughout each layer of the atmosphere \citep{Nicholls2024MNRAS}. Our strictly energy conserving formalism permits the formation of convectively-stable layers within the atmosphere, yielding realistic cooling rates and climates \citep{Selsis2023Nature, Guillot2010AA, Cmiel2025PSJ}. Spectroscopic radiation fluxes are calculated with the \textsc{socrates} radiative transfer code \citep{Edwards1996QJRMS, Manners2024SOCRATES, Sergeev2023GMD}, under the two-stream and plane-parallel approximations. Opacities are combined using a correlated-$k$ method \citep{Lacis1991JGR}, including Rayleigh scattering, but neglecting clouds or other aerosols \citep{Madhusudhan2016SSRev, Janssen2026arXiv}. Atmospheric convection is parametrised using mixing length theory  under the Schwarzschild criterion \citep{Joyce2023Galaxies, Marley2015ARAA, Vitense1953ZfA, Hogstrom1988BLM}.

The incoming stellar radiation represents the top boundary condition for our radiative-convective calculations, and is obtained from the \textsc{mors} stellar evolution model in parallel to the X-ray and ultraviolet (XUV) irradiation fluxes \citep{Johnstone2021AA}. Stellar rotational evolution is treated using a two-shell approach; calculated angular momentum exchange within the star and with the stellar wind provide XUV fluxes through empirically-derived scaling relations, as functions of stellar age \citep{Spada2013ApJ, Brun2017LRivSP}.

\subsection{Hydrodynamic escape model}
\label{sec:methods_escape}

We model atmospheric mass loss using a parametrisation of hydrodynamic energy-limited escape \citep[e.g.,][]{Watson1981Icarus, Lammer2003ApJL}, which dominates on close-in planets during the early high-XUV phase of stellar evolution \citep[e.g.,][]{Jackson2012MNRAS, Pezzotti2021AA, Attia2025AA}. In this formulation, stellar XUV flux absorbed at the top of the atmosphere heats the gas, driving a bulk outflow at a rate $\dot{M}_\mathrm{bulk} = \eta \pi R_\mathrm{XUV}^3 F_\mathrm{XUV} / (G M_\mathrm{p})$,
where $\eta$ is the escape efficiency, i.e., the fraction of absorbed XUV energy lifting the escaping gas, $R_\mathrm{XUV}$ is the radius at which the atmosphere becomes optically thick to XUV radiation, $F_\mathrm{XUV}$ is the incident stellar XUV flux, $G$ is the gravitational constant, and $M_\mathrm{p}$ is the planet mass. $R_\mathrm{XUV}$ is determined self-consistently from the atmospheric structure computed by the climate module \textsc{agni}, evaluated at a reference pressure $P_\mathrm{XUV}$ of 20\,mbar \citep{Lehmer2017ApJ, Lopez2014ApJ}. The time-dependent $F_\mathrm{XUV}$ is provided by the \textsc{mors} stellar evolution model \citep{Johnstone2021AA} using rotation-activity evolution tracks from \citet{Spada2013ApJ}.

We treat the escaping outflow as compositionally unfractionating; each volatile element $e$ escapes at a rate proportional to its atmospheric mass fraction, $\dot{M}_e = (M_e/M^\mathrm{atm}) \times \dot{M}_\mathrm{bulk}$, where $M_e$ is the mass of element $e$ in the atmosphere and $M^\mathrm{atm}$ is the total atmosphere mass (Section~\ref{sec:methods_atmos}). Although the escape process itself does not preferentially remove any species, it nonetheless fractionates the \emph{bulk} planetary volatile inventory because the escaping composition reflects the outgassed atmospheric mixture, which differs from the total planetary (mantle plus atmosphere) inventory \citep{Nicholls2026NatAs}. Our escape treatment thus does not preferentially remove or retain species within the outflow itself, but rather includes the bulk \emph{atmospheric} elemental inventory in the  outflow. Therefore, low solubility elements, like carbon \citep{Bower2022PSJ}, are preferentially concentrated in the atmosphere and thus lost to space, and this is taken into account in molecular abundances calculations (e.g.\,CO). Strong coupling between redox-controlled atmospheric speciation and escape is the mechanism that drives the compositional shift central to the reflation effect. We note that species-dependent fractionation in the upper-atmosphere outflow \citep[e.g.,][]{Schulik2023MNRAS,Cherubim2024ApJ,Cherubim2025ApJ,KrissansenTotton2024NatCo,Valatsou2026arXiv}, not modelled here, may further modulate the MMW evolution and the magnitude of reflation.

At each simulation time step, the escape, outgassing, and climate calculations are coupled. After updating $F_\mathrm{XUV}$ as a function of stellar age, initialised at 100\,Myr, we compute the bulk mass-loss rate and distribute it across the CHNOS elements. We then subtract the integrated per-element mass loss $\dot{M}_e \times\mathrm{d}t$ from the bulk volatile inventory. The outgassing module (\textsc{calliope}) then speciates the remaining volatile masses according to the chemistry, solubilities, and the mantle redox state. Finally for a given iteration, the atmosphere thermal structure and the new $R_\mathrm{XUV}$ are recalculated. This coherent coupling ensures that changes in atmospheric composition and structure interact with the outgassing and the escape, capturing non-linear feedback between interior geochemistry and atmospheric evolution.

\subsection{Simulation ensemble parameters}

Ultra-short period rocky planets orbit very close to their stars ($<$0.5~AU) with orbital periods on the order of 1~day. Therefore, due to their close-in orbits, they are expected to be tidally locked into axial-orbital rotation synchrony and subject to extreme instellation fluxes on the order of $10^3$ times Earth's. Due to their extreme irradiation conditions, ultra-short period planets can readily retain a molten state, and are likely to lose their atmospheres due to stripping from their host star. However, recent observational studies have shown some of these planets to be inconsistent with bare rock surfaces \citep{Zieba2022AA, Hu2024Natur, August2025AA, Monaghan2025AJ, Teske2025ApJL, ParkCoy2026arXiv}. 

In this study, our original goal was to understand under which initial conditions and physical regimes ultra-short period planets retain their atmospheres. The simulations shown in this work are inspired by the physical characteristics of to the ultra-short period super-Earth TOI-500\,b: a $1.42M_{\oplus}$ planet orbiting a K6V star ($M_\star=0.74M_{\odot}$) at 0.01\,AU \citep{Serrano2022NatAs, Giacalone2022AJ}. However, we perform a suite of simulations to test the robustness of the mechanism across a wide parameter range. To initialise evolutionary simulations of this planet we consider a range of potential post-formation scenarios with hydrogen inventories $H_\mathrm{ocean}=[1,5,10,20]$ (in mass units of Earth's oceanic hydrogen budget) and mantle redox states proxied by $f\ch{O2}$ ($ \Delta\text{IW}=[0,+2,+4]$). Given the uncertainties in the physics and rate of escape \citep{Owen2019AREPS,Gronoff2020JGRA,Kubyshkina2026SSRv}, we vary the energy efficiency term $\eta=[10^{-2},10^{-3},10^{-4},10^{-5}]$, and simultaneously explore a range of orbital semi-major axes $a = [0.1,0.05,0.02,0.01]\mathrm{AU}$. Our parameter space allows exploration of the long-term evolution of a ultra-short period exoplanets (like TOI-500\,b) spanning different instellation and geochemical environments, escape regimes and chemical inventories. All of these parameters factor into atmospheric evolution. We discuss the implications and limitations of these assumptions in Sect.~\ref{sec:limitations}.

\section{Results}
\label{sec:results}

\begin{figure*}[htb]
    \centering
    \includegraphics[width=0.9\textwidth]{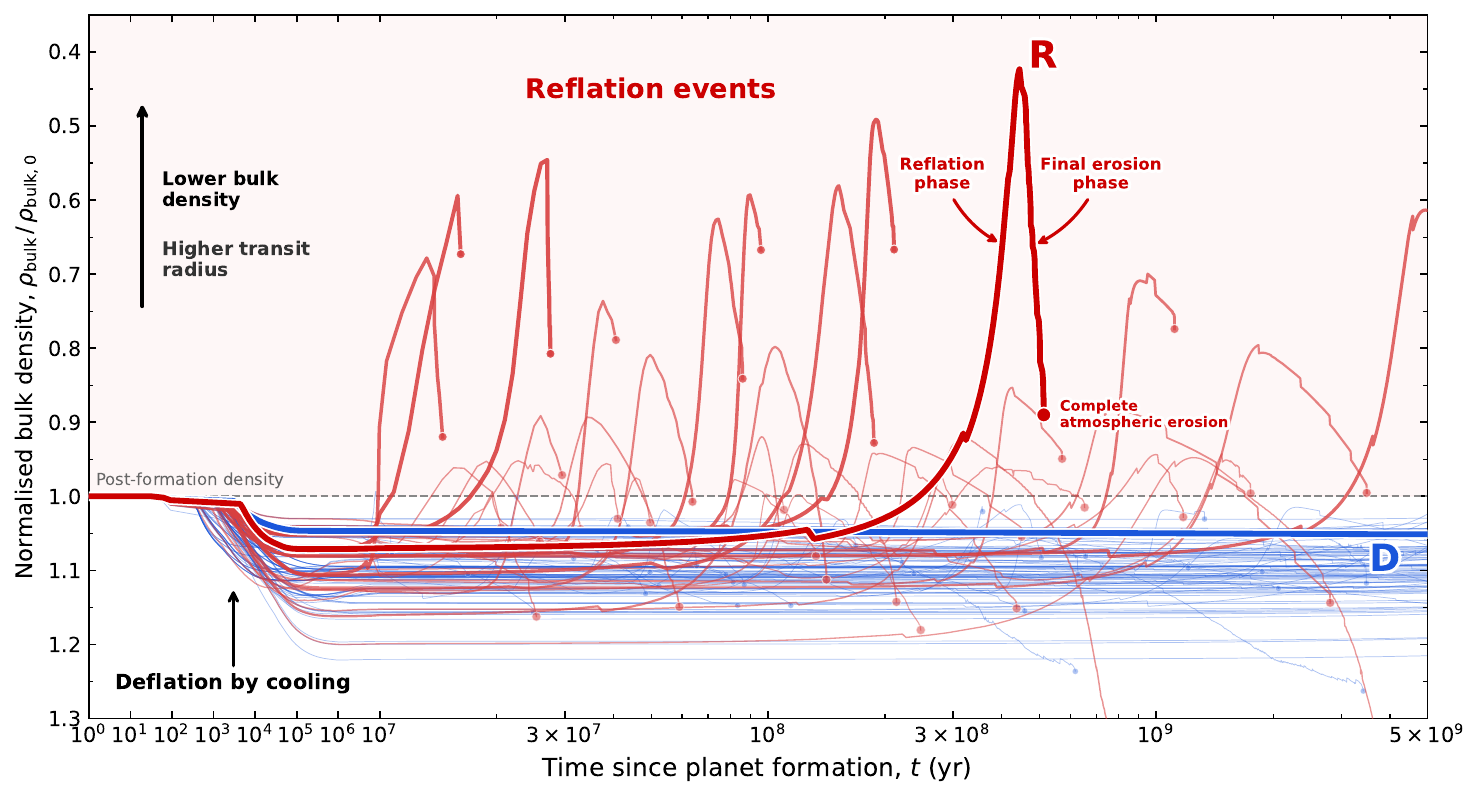}
    \caption{Reflation events (red) occur across a range of ages and with different magnitudes across the evolution of irradiated super-Earths. The plot shows the normalised bulk density evolution of individual simulations, versus the time elapsed since planet formation. The red lines indicate cases where a reflation event occurs, while blue lines indicate those where it does not occur. The evolution tracks that end with a dot indicate the planet has reached an airless state and complete atmospheric erosion by hydrodynamic escape has taken place. The evolution of the two select cases for Figure \ref{fig:2} are highlighted and denoted with a \textbf{R} and \textbf{D}. These cases illustrate in more detail a representative \emph{reflation} (red) and \emph{deflation} case (blue).}
    \label{fig:2}
\end{figure*}

\begin{figure}[htb]
    \centering
    \includegraphics[width=0.48\textwidth]{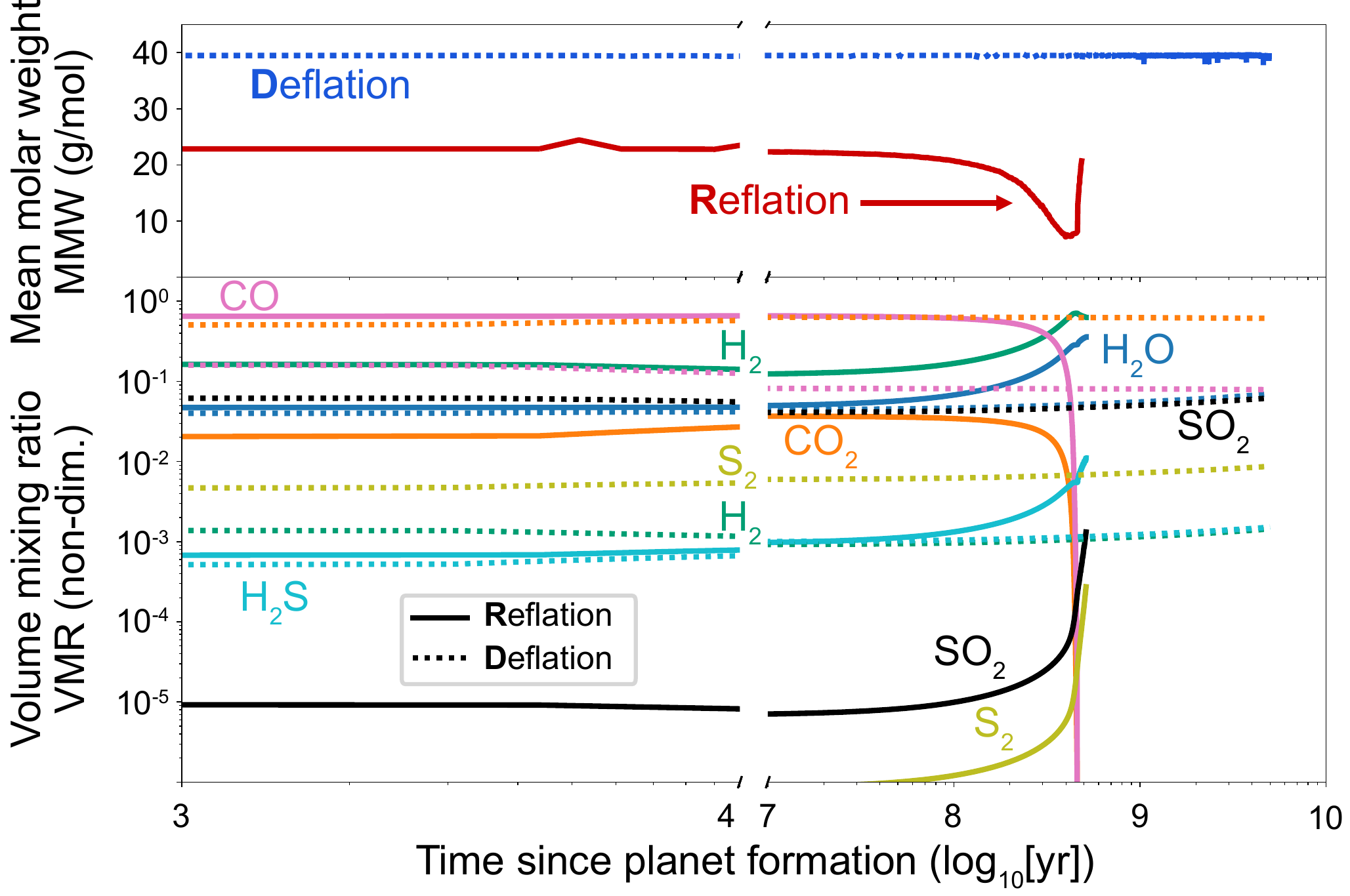}
    \caption{Evolution of atmospheric composition for the labelled reflation (solid lines) and deflation (dashed lines) cases from Figure \ref{fig:2} (\textbf{R} and \textbf{D}). \emph{Top panel}: evolution of the atmospheric mean molar weight. \emph{Bottom panel}: the volume mixing ratio (VMR) of the dominant molecules in the planets' atmospheres. The reflation simulation loses its primary atmospheric gas (CO, colored pink) through hydrodynamic escape. Until this point, the H molecules of the reflation case are dominantly dissolved as H$_2$O in the underlying magma ocean, and outgassed to the atmosphere as the CO-dominated envelope escapes to space. This transition from C- to H-dominated atmosphere re-inflates the atmosphere (through the decrease in MMW) and reduces the observed bulk density of the planet transiently by up to 60\%.}
    \label{fig:3}
\end{figure}

\begin{figure*}[htb]
    \centering
    \includegraphics[width=0.95\textwidth]{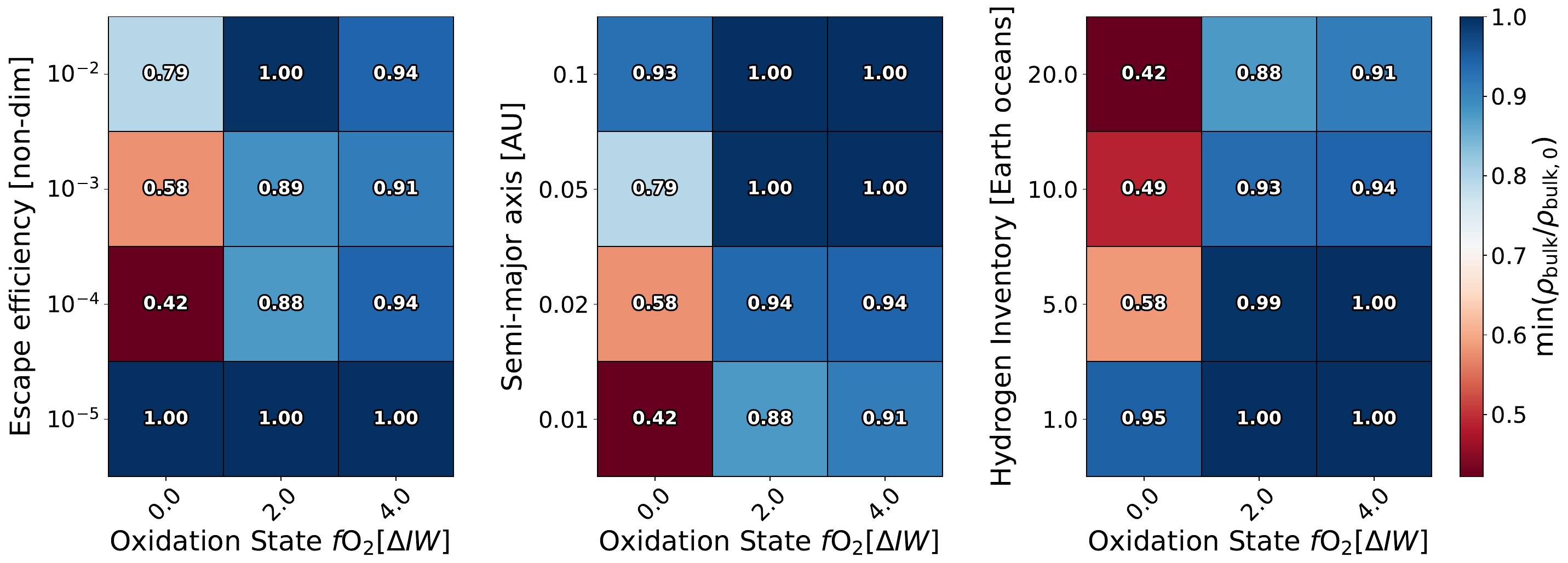}
    \caption{Sensitivity of the reflation mechanism on mantle oxidation state, atmospheric escape efficiency, planetary semi-major axis, and initial hydrogen inventory. Reflation events occur most strongly at mantle oxidation state $f$O$_2$ $\sim$ IW, escape efficiency $\eta \sim$ $10^{-4}$ to $10^{-3}$, semi-major axis $a \lesssim$ 0.05 au, and initial hydrogen inventories $\gtrsim$ 5 Earth oceans. The sensitivity matrices display the minimum normalised bulk density that is achieved across three projections of the simulation ensemble. We show a range of redox states (\textbf{x-axis}) and respective simulation parameters (\textbf{y-axis}). For example, among all simulations with a mantle redox state of IW+0 and semi-major axis of 0.02 au in the middle panel (which includes all variability of escape efficiency and hydrogen inventory), the most reflated case reached an under-density of $\rho_{\mathrm{bulk}}$ $58\%$ relative to its post-formation bulk density $\rho_{\mathrm{bulk,0}}$.}
    \label{fig:4}
\end{figure*}

Figure \ref{fig:2} presents an ensemble of planetary evolution pathways, plotting the bulk density $\rho_\mathrm{bulk}$ over time (normalised to its initial value in each case, $\rho_\mathrm{bulk,0}$) . The lines are colored based on whether reflation events emerge before complete atmospheric erosion (red), or if no reflation occurs and the planet either loses its atmosphere through monotonic atmospheric loss, and bulk density increase, or retains its secondary atmosphere for a minimum of 5 Gyr (blue), which is the observed age of TOI-500, our fiducial model system. The y-axis of Figure \ref{fig:2} is reversed such that a decrease in bulk density (i.e., atmosphere inflation) is shown as an upward spike. We categorize a simulation as `reflation' in case it re-inflates to such a degree that it (again) reaches its initial bulk density some time after its initial deflation. The deflation phase is largely driven by cooling post-formation \citep{Nicholls2026NatAs, Tang2024ApJ, calder_most_2025}. Reflation events traverse density signatures from barely below the post-formation bulk density down to $\sim$40\% of the initial density. Across our models, reflation is found to occur at a range of planet ages, with the first events occurring shortly after planet formation and disk dissipation ($\sim 100$\,Myr), while some cases as late as 5\,Gyr, the TOI-500 system's estimated current age \citep{Serrano2022NatAs}. The magnitude, timing, and likelihood to occur depend on formation conditions and interior geochemistry, which we now explore.

To illustrate the mechanism behind these reflation events, we highlight two representative simulations from Figure \ref{fig:2}. They represent the two principal outcomes of our simulations for these planets: a geochemically-reduced case where reflation occurs before complete atmospheric erosion (denoted with \textbf{R} for \emph{R}eflation), and an oxidised case where the planet retains its atmosphere's heavier composition indefinitely (denoted with \textbf{D} for \emph{D}eflation). Figure \ref{fig:3} displays the chemical evolution of the atmosphere of the two representative cases, showcasing the characteristic behaviour of atmospheric compounds in the reflating (solid line, \textbf{R}) or deflating (dashed lines, \textbf{D}) cases. The top panel plots the atmospheric mean molecular weight (MMW) of the atmospheres over time, where the reflating planet's MMW decreases strongly during the reflation event, while the MMW of the deflating case remains approximately constant throughout. This divergence is caused by the loss of the initial CO-dominated secondary atmosphere in the reducing case. \textbf{R}'s atmosphere is dominated by CO because outgassing at geochemical conditions comparable to the IW buffer favours CO production. In this stage, H is dominantly stored in the underlying magma ocean in the form of H$_2$O, and only a minor part of it is outgassed as gas in the atmosphere. \textbf{R}'s reflation event is taking place at around $3 \times 10^8$ yr, when the initial CO-dominated atmosphere is eroded to space by hydrodynamic escape. This decreases the surface pressure, which enables the outgassing of hydrogen, which splits into H$_2$ and H$_2$O according to solubility and gas phase chemistry at these conditions \citep{Sossi2023EPSL}. Over the course of the loss of CO-dominated atmospheres this process transiently replaces the secondary atmosphere with a lighter envelope where hydrogen dominates. This is shown in the bottom panel of Figure \ref{fig:3}. Accordingly, the MMW of the atmosphere decreases and the scale height increases until the majority of the hydrogen is also eroded to space. In the final stage of the reflation event, the remaining S-dominated species degas to the atmosphere, which produces a turnover towards higher atmospheric MMWs. Sulfur-bearing gases are highly soluble under reducing conditions \citep{Gaillard2022EPSL}, however, the strong decrease in total atmospheric pressure releases them into the atmosphere \citep{Bower2022PSJ}. This release happens at $\sim5 \times 10^8$ yr in the specific case of the \textbf{R} simulation, for which the reflation event lasts $\sim$200 Myr. During the course of the \emph{reflation} event the melt fraction does not change substantially and remains at $\gtrsim 90\%$ throughout the reflation simulation and at 100 \% throughout the \emph{deflation} simulation in Figure \ref{fig:3}. In the \emph{reflation} case, the mantle starts to crystallize once the atmosphere starts deflating after the peak of the reflation event. Figure \ref{fig:2} shows that these events have a wide range of onset times (from $\sim 10^7$ yr to several $\sim10^9 yr$) and durations. Each duration is roughly on the same order of magnitude as the onset time after planet formation. 

In the oxidised scenario of Figure \ref{fig:3}, the atmosphere transitions over time from a $\text{CO}_2$-dominated to a composition that is becoming slightly enriched in H$_2$O, but survives for a substantially longer time. The MMW of the oxidised atmosphere is buffered by the continuous degassing of $\text{SO}_2$, which is soluble in the magma ocean at reduced conditions, but \emph{not} at oxidized mantle redox state, and hence outgasses to the atmosphere throughout \citep{Shorttle2024ApJL,Nicholls2026NatAs}. All oxidised scenarios behave similarly, those with and without eventual complete atmospheric erosion, where the initially $\text{CO}_2$- and $\text{SO}_2$-dominated envelope  becomes more enriched in H$_2$O over time. Critically, in all of these cases, the atmosphere remains enriched in heavy species that are dominated by oxygen, and hence the MMW of the atmosphere remains roughly constant, and no late re-inflation of the atmosphere takes place.

Figure \ref{fig:4} presents a parameter study of the sensitivity of reflation events on escape efficiency, semi-major axis, and initial hydrogen inventory of the super-Earths. Each colored square in Figure \ref{fig:4} displays the lowest value of the normalised bulk density (i.e., magnitude of reflation) at various mantle redox states, atmospheric escape efficiencies, orbital semi-major axes, and initial hydrogen inventories from the explored simulation parameter space. The mantle oxidation state (as a proxy for internal geochemistry) is the key controlling factor in regulating the occurrence and magnitude of reflation events. All dominant reflation peaks are observed in the leftmost column of each Figure~\ref{fig:4} panel, where the redox state is equal to the iron-w\"ustite buffer. In general, any value of the minimum normalised bulk density achieving $\le1.0$ represents a parameter regime where reflation occurs for at least one scenario inflated above its post-formation bulk density.

The leftmost panel of Figure \ref{fig:4} indicates that the strongest reflation is achieved at intermediate escape efficiencies ($\eta=10^{-3}$~and~$10^{-4}$) because higher efficiencies generate a moderate peak while mass loss processes become insignificant at lower efficiencies. At very high escape efficiencies the atmosphere is lost too rapidly. The central panel of Figure \ref{fig:4} shows that reflation events are sensitive to the instellation flux, where reflation is occurring most strongly for the innermost orbit of the most reduced cases. Reflation becomes significant below $\approx$0.05 au and farther orbits lead to smaller to no discernible reflation events. Finally, the rightmost panel of Figure \ref{fig:4} highlights the importance of the initial hydrogen inventory in fuelling reflation events. Planets with more hydrogen available in their atmosphere and mantle spend more time in the reflation phase before the atmospheric erosion causes the atmosphere to deflate. This means that a minimum initial inventory of hydrogen must be present from planet formation. Reflation is significant above a hydrogen budget greater than 5 Earth ocean equivalents. For reference, the combined H budget of the Earth's mantle plus atmosphere is $\sim$2-10 Earth oceans \citep{Peslier2017SSRv}.

\section{Discussion}

\subsection{Reflation as observable signature of interior geochemistry}

The reflation mechanism demonstrated through our simulations suggests a direct and potentially observable link between a rocky planet's interior geochemistry, volatile budget, and the time evolution of its observed transit radius and bulk density. In conventional models of atmospheric escape, mass loss monotonically strips the envelope and increases the bulk density over time \citep{Owen2017ApJ, Gupta2019MNRAS, Rogers2025arXivb}. Reflation breaks this monotonic trend by incorporating the effects of mantle degassing. For reduced scenarios, escape-driven depletion of the initially CO-dominated atmospheres is counteracted by outgassing of progressively lighter species (H$_2$O, H$_2$) from the melt, because they are more soluble \citep{Dorn2021ApJL, Bower2022PSJ, Sossi2023EPSL, Suer2023FrEaS}. Later-stage outgassing of lighter species lowers the MMW and transiently inflates the atmosphere, even as total volatile mass decreases. Non-monotonic density evolution as a direct result from interior-atmosphere feedback is a qualitatively novel prediction that cannot arise in models treating atmospheric composition as fixed or neglecting the interior volatile reservoir \citep{Lichtenberg2025TrGeo}.

The physical origin of reflation lies in the redox-dependent competition between volatile solubility and outgassing speciation. At low oxygen fugacity (near the iron--w\"ustite buffer), carbon is outgassed primarily as CO ($\mu \approx 28$~g~mol$^{-1}$), which dominates the initial atmospheric mass. All S compounds remain dissolved in the magma ocean at these redox conditions \citep{Gaillard2022EPSL,Suer2023FrEaS}. Hydrogen is stored in the melt as dissolved H$_2$O; as escape strips the CO-rich atmosphere, the decreasing surface pressure shifts the solubility equilibrium, degassing this water into the gas phase. Once in the atmosphere, equilibrium chemistry at the prevailing reduced conditions converts much of the H$_2$O into H$_2$ ($\mu \approx 2$~g~mol$^{-1}$) \citep{Sossi2023EPSL, Thompson2025ChemGeo}, progressively replacing heavy carbon-bearing species with light hydrogen-bearing ones. The resulting compositional shift lowers the MMW and inflates the scale height. In the final phase of a reflation event, the declining pressure at the surface also degasses sulfur species, such that the last atmosphere before complete desiccation is S-rich. In contrast, oxidized mantles initially outgas a CO$_2$-dominated atmosphere ($\mu \approx 44$~g~mol$^{-1}$), including contributions from CO ($\mu \approx 28$~g~mol$^{-1}$) and $\text{SO}_2$ ($\mu \approx 64$~g~mol$^{-1}$), which increases the MMW, decreases the scale height, and stabilises the atmosphere against efficient hydrodynamic loss. H$_2$O in the oxidised case is degassed only much later as the surface pressure drops further. Because these oxidised gases are heavy, their delayed transition does not produce a comparable MMW decrease, and the atmosphere deflates monotonically.

This redox sensitivity, accessible via existing planetary-evolution models, makes reflation a possible observational signature of the mantle geochemistry occurring within super-Earth exoplanets. Sub-Neptunes and larger planets retain thick primary envelopes of H$_2$/He, whose low MMW already dominates the atmospheric scale height; any secondary outgassing signal is diluted by the primordial gas, and the stronger surface gravity suppresses the relative change in scale height from compositional shifts. Reflation requires that a secondary, outgassed atmosphere constitutes the bulk of the envelope, so redox-driven changes in speciation can materially alter the MMW. This confines the reflation mechanism to super-Earths that have already lost (or never accreted) a significant primary atmosphere, yet still host a molten, volatile-bearing interior. The indications of heat-redistribution signatures on highly irradiated super-Earths \citep{Zieba2022AA, Hu2024Natur, August2025AA, Monaghan2025AJ, Teske2025ApJL, ParkCoy2026arXiv} increasingly favour the existence of such types of heavy secondary atmospheres. Measurements of low bulk densities (i.e., an anomalously extended atmosphere) on highly irradiated super-Earths, inconsistent with bare-rock or monotonic-stripping scenarios, would point to an actively degassing, geochemically-reduced interior. The magnitude, timing, and duration of reflation depend on the oxygen fugacity, irradiation environment, initial volatile endowment, and escape efficiency (Sect.~\ref{sec:results}). Simultaneous constraints on density and atmospheric composition from transit and emission spectroscopy could, in principle, narrow the range of plausible interior redox states (Sect.~\ref{sec:obs}). The search and identification of low-density reflated super-Earth planets presents a valuable opportunity for precisely probing exoplanetary subsurface geochemistry through atmospheric markers. Deep characterisation of mantle redox complements the existing approach of bulk-density constraints, which are otherwise degenerate between composition and thermal state \citep{Seager2007ApJ, Rogers2010ApJ, Zeng2016ApJ, Dorn2021ApJL,Boer2025ApJ}.

\subsection{Dependence on primary parameters}

Reflation events are observed only in a subset of our simulations. Event occurrence depends most critically on a planet's mantle redox state, while reflation magnitude varies according to a planet's initial volatile inventory, instellation, and escape efficiency. The strongest peak is observed for a reduced case (where reduced refers to an oxidation state of $\Delta \mathrm{IW} = 0$) with a high hydrogen content ($H_\mathrm{ocean} = 20$), a closer-in orbit ($a=0.01$ AU), and a low escape efficiency ($\eta=10^{-4}$). As shown in Figure \ref{fig:4}, the reflation mechanism is most pronounced for simulations falling in the reduced region of the investigated parameter space, corresponding to oxygen fugacities $f$O$_2$ between  $\Delta$IW$\pm$0 and $\Delta$IW+2. Within this regime, the atmosphere is initially CO dominated, which enables the reflation mechanism as discussed above,  In contrast, planets with more oxidized interiors ($\Delta$IW $\gtrsim$ +2) do not exhibit significant inflation regardless of escape efficiency, orbital distance, or initial hydrogen inventory. For these oxidized mantles, the outgassed composition remains dominated by heavy molecules (mainly CO$_2$) throughout the evolution, and the small changes in mean molecular weight are insufficient to counterbalance the mass loss. The resulting atmospheric evolution is monotonic deflation, with bulk density increasing steadily over time.

Beyond redox state, other parameters modulate the timing and magnitude of reflation. Higher initial hydrogen inventories provide a larger reservoir of hydrogen-rich volatiles that can sustain the late-stage outgassing of H$_2$O and H$_2$, both by directly contributing water and by enabling more vigorous melt production through lower solidus temperatures. Furthermore, the higher hydrogen content leads to stronger peaks as the atmosphere continues inflating, and thus increasing its $F_\mathrm{XUV}$ absorbing radius, as long as it has enough hydrogen to drive this effect. More efficient escape ($\eta$ closer to unity) accelerates the removal of the initial CO-rich atmosphere, hastening the transition to the hydrogen-dominated outgassing regime. Shorter orbital distances increase the incident XUV flux and thus the escape rate, which similarly speeds up the compositional shift. The interplay between these parameters defines a finite window in parameter space where reflation occurs: the planet must have enough hydrogen to supply the later outgassing phase, sufficient escape to remove the initial heavy atmosphere but not too efficient as to lose most of its hydrogen content too fast impeding the atmospheric reflation, and a redox state that allows the outgassed composition to change substantially over the course of the evolution.

Candidate targets that appear to fall within, or near, this parameter space by having close-in orbits and appearing to be under-dense, indicative of significant hydrogen inventories, are already known in the literature. See for example the Kepler-11 system \citep{lissauer_all_2013, ikoma_situ_2012} where follow-up modeling quantified radius inflation due to H/He \citep{wolfgang2015}; Kepler-51 system \citep{masuda2014, libby-roberts2020, masuda2024}; GJ 1214b \citep{charbonneau2009, kempton2023}; 55 Cancri e \citep{hu_secondary_2024}; and TOI-561b \citep{Teske2025ApJL}. We propose such targets as promising candidates according to our theoretical analysis.

\subsection{Uncertainties and limitations}
\label{sec:limitations}

PROTEUS incorporates the key physical and chemical interactions linking planetary interiors and atmospheres. However, as with any model, the model incorporates substantial approximations, which may merit special attention in future work. In this section we discuss the major uncertainties and model-specific considerations that may affect the reflation mechanism.

An important consideration, in the context of the reflation effect, is that our atmospheres are treated with vertically well-mixed chemical compositions determined at the surface by their partial pressures from the outgassing calculation. In reality, the combined effects of thermochemistry and various disequilibrium processes (e.g. photochemistry and diffusion kinetics) will modulate gas mixing ratios and influence the transport of radiation through the atmosphere \citep{Hu2012ApJ, Hu2013ApJ, Madhusudhan2016SSRev, Nicholls2023MNRAS, Werlen2025ApJL}. However, chemical reprocessing of atoms into various gas-phase species will only negligibly impact the atmospheric height structure (and thus, the bulk planet density) because the atmospheric mean molecular weight is largely a function of atmospheric \textit{elemental} mixing ratios. Similarly, we treat the escaping outflow's \textit{elemental} composition as equivalent to that degassed equilibrium at the surface. Elemental inhomogeneity may be induced by cold-trapping in colder environments than considered here \citep{Habib2024ApJ, Pierrehumbert2010Book} or by de-mixing by chemical immiscibility, which is applicable to the sub-Neptune and gas giant regimes \citep{Piaulet2025arXiv, Gupta2025ApJL, Rogers2025arXiva}. 

A related uncertainty concerns the choice of escape efficiency $\eta$. The canonical value of $\sim$ 10\% commonly adopted in energy-limited mass-loss calculations originates largely from studies of hot Jupiters and H$_2$/He-dominated primary envelopes \citep[e.g.,][]{MurrayClay2009ApJ}, where stellar photons couple efficiently into a deeply extended, low MMW atmosphere. Radiation-hydrodynamic (RHD) simulations of water-rich atmospheres on rocky planets \citep{Owen2016ApJ}, as applied to the TRAPPIST-1 system by \citet{Bolmont2017MNRAS}, indicate instead that $\eta$ is not constant but depends sensitively on the incident XUV flux, spanning $\sim$$10^{-3}$ to $\sim$$10^{-1}$, with the lowest values reached at the high XUV luminosities characteristic of young, close-in systems. This decrease with increasing flux reflects the fact that Ly$\alpha$ and molecular line cooling divert a growing fraction of the absorbed energy from mechanical work on the gas \citep{MurrayClay2009ApJ, Owen2016ApJ}. Our adopted $\eta$ values, which are held constant throughout simulations, fall within the physically motivated range for high XUV regimes where line cooling is also taken into consideration as further suppressing escape \citep{nakayama_survival_2022, Yoshida2024PEPS, Chatterjee2026ApJ} These $\eta$ values are thus justifiably lower than those typically assumed for gas-rich envelopes. For the high MMW, CO/CO$_2$-rich secondary atmospheres central to the reflation mechanism, no dedicated RHD simulations quantifying escape efficiency currently exist, to the best of our knowledge. The compressed scale height, smaller XUV-absorbing  cross-section, and more efficient infrared cooling by polyatomic molecules \citep{wordsworth_fermi_2024} likely push $\eta$ even lower than the \citet{Bolmont2017MNRAS} range, though the exact values remain unknown. Comparable conclusions have recently been reached by \citet{Yoshida2024PEPS, Yoshida2025AA,Ji2025ApJ,Chatterjee2026ApJ}. Importantly, lower efficiencies do not suppress reflation but extend its characteristic timescales, which would broaden the observational window for detecting transient underdensities on young super-Earths.

We treat the escaping outflow as compositionally unfractionating, and assess the validity of this assumption by comparing our energy-limited mass flux $f = \eta F_\mathrm{XUV}/(4V_\mathrm{pot})$ to the critical crossover flux $f_c = b\,x_1\,(m_2 - m_1)/H_1 $. Above this crossover flux, heavy species are dragged along with the escaping gas in the bulk, in a non-fractionating regime \citep{Hunten1987Icarus}. We evaluate this following the implementations of \citet{Cherubim2024ApJ} and \citet{Wordsworth2018AJ}, adopting parameters representative of our simulations together with binary diffusion coefficients from \citet{Wordsworth2018AJ}. We find that for an inflated H\textsubscript{2}-dominated atmosphere during the XUV-saturated stellar phase, at $a\lesssim 0.05$ au and $\eta \gtrsim 10^{-4}$, the escape flux exceeds the critical flux by one to three orders of magnitude ($f/f_c\sim2\, \text{to}\ 760$). Our assumption of unfractionating escape is strongly supported in this regime. Diffusive fractionation ($f\leq f_c$) is instead confined to the weak-escape regime of our parameter space (low efficiency, wide orbits, and the post-saturation XUV phase), where mass loss is slow and reflation is correspondingly weak or absent (Fig.~\ref{fig:4}). The crossover formalism applies generally to any light major constituent dragging a heavy species \citep{Wordsworth2018AJ}, we evaluated it here for the H\textsubscript{2}-dominated inflated phase in reflation events.
Photochemistry will dissociate the dominant carbon species (CO, and later CO\textsubscript{2}) in the upper atmosphere, so the relevant question is not whether carbon and oxygen escape as molecules or as atoms, but whether the escape flux exceeds the crossover flux for the masses involved \citep[a species-resolved problem addressed by multispecies hydrodynamic escape models as][]{Schulik2023MNRAS}. In the vigorous-escape regime that drives reflation the flux lies one to three orders of magnitude above crossover, placing the crossover mass above that of CO (28 g/mol) and hence well above the atomic masses of carbon (12 g/mol) and oxygen (16 g/mol). Because the photodissociation products are lighter than their parent molecules, they remain entrained in the bulk outflow wherever the parent molecules are, and over a strictly larger region of parameter space. Dissociation therefore relaxes the crossover condition rather than tightening it. Elemental separation of carbon and oxygen becomes possible only in the weak-escape corner of our parameter space (wide orbits, low efficiency, and the post-saturation XUV phase), where the crossover mass drops below 16 g/mol, but there mass loss is slow and reflation is itself weak or absent (Fig.~\ref{fig:4}), consistent with the transition from bulk-dragged to diffusively fractionating escape found in hydrodynamic simulations by \citet{schulik2025}.

Crucially, the reflation signature originates from the interior-atmosphere partitioning of preferentially-dissolved H2O and its conversion to H2 as the surface pressure decreases, rather than from escape preferentially retaining hydrogen. Upper-atmosphere fractionation would therefore modulate the magnitude and timing of reflation rather than suppress the mechanism itself, consistent with the conceptual scope of this study. A more self-consistent treatment requires coupling diffusive fractionation directly into the volatile speciation–solubility solver, enabled by various transport processes through the atmosphere. This mechanism is addressed by current open-source fractionation codes \citep[e.g., IsoFATE coupled to the magma ocean speciation solver Atmodeller][]{Cherubim2025ApJ, Bower2025ApJ}, applicable to the  high-MMW, magma-ocean-coupled regime. Incorporating this effect in our simulations would likely affect the magnitude of reflation, without erasing it. Quantifying its effect is left for future work.

Interior-atmosphere partitioning calculations have largely assumed that volatiles dissolved within ascending parcels of silicate melt become saturated near the magma ocean surface, such that chemical reprocessing conditions are representative of bubble exsolution conditions \citep{Schaefer2018ChemGeo, Lichtenberg2023ASPC, Salvador2023SSRv}. Laboratory and numerical studies suggest that magma ocean degassing processes do not necessarily occur at equilibrium \citep{walbecq_the_2025}, subject to turbulent fluid dynamics and bubble formation microphysics \citep{Pioli2012JVGR, Dasgupta2013RMG, Gardner2023PSJ, Salvador2023Icar}. Additionally, hydrogen may partition into the \textit{solidified} deep mantle via entrained magma pockets \citep{HierMajumder2017JGRP, Sim2024JGRP}. Disequilibrium outgassing and inhomogeneous chemical stratification of volatiles throughout the deep mantle both allow greater volatile storage capacities; effectively degenerate with the range of initial volatile inventories and escape efficiencies considered here \citep{Nicholls2026NatAs}.

We have adopted a range of potential $f\ch{O2}$ offsets relative to the iron--w\"ustite buffer, which are held fixed during each simulation. Our approach intuitively parametrises multiple geodynamic and geochemical processes that collectively determine the redox state and oxygen fugacity throughout the mantle of Earth \citep{Frost2008AREPS,2025NRvEE...6..728C} and super-Earth exoplanets \citep{,Wordsworth2022ARAA,Lichtenberg2025TrGeo,Lichtenberg2025Science}; e.g., core segregation, the rainout of metal droplets from the mantle, and phase stabilisation by electron spin alignment \citep{Lichtenberg2023ASPC,Lichtenberg2025TrGeo,zhang_disproporti_2014}. We emphasise that the mantle oxygen fugacity $f\ch{O2}$ offset relative to the IW buffer is held fixed throughout each simulation \citep{ONeill2002ChemGeo}, so the reflation events identified here reflect the planet's instantaneous geochemical state rather than a self-consistently evolving redox history. A shift towards more oxidised mantles at peak hydrogen loss would push an initially reducing mantle towards a more oxidised state \citep{kasting1993, katyal2020}, however reflation is contingent on the planet residing in a sufficiently reduced state ($f\ch{O2} \lesssim \Delta IW+2$) at the moment its initial CO-dominated atmosphere is stripped. Provided this condition is met within the relevant evolutionary window, the qualitative compositional transition that drives reflation is preserved. A self-consistent treatment of mantle redox evolution would thus refine the time and magnitude of the onsets (Figure \ref{fig:2}) and the quantitative boundaries of the reflation regime (Figure \ref{fig:4}), without removing the mechanism itself. Dissolved volatiles further complicate the geochemistry by directly exchanging electrons with mantle phases to additionally modulate the redox state. Although subject to these empirical and conceptual uncertainties, prior theoretical studies have suggested that hydrogen escape could \textit{oxidise} mantle chemistry over time \citep{Catling2001Sci,Wordsworth2014ApJ, Hirschmann2023EPSL, Cherubim2025ApJ}.

Finally, each of our simulations adopt a planetary interior structure solution which is held fixed over time. This neglects the impact of deep mantle solidification on the density of these layers, but is largely negligible here because our simulations remain in the (semi-)molten regime \citep{Wolf2018PEPI, Boley2023ApJ}. Similarly, the atmosphere's mass would marginally compress the mantle-building material \citep{Hirose2021NatREE}. Future avenues of research may consider the reflation effect in the sub-Neptune regime, where these structural considerations become important \citep{ Unterborn2023ApJ, Dorn2015AA}.

\subsection{Opportunities for observability}
\label{sec:obs}

The concept of reflation highlighted here illustrates the importance of transient evolutionary processes in shaping the observed population of exoplanets. When observable, such processes can be used as probes for atmospheric evolution and calibrate our picture of planetary formation. Reflation events would leave observable imprints on bulk properties, such as radius and density, with observable drops between few percent to over $60\%$ compared to standard evolutionary models. Further, the compositional shift in the atmospheric composition that drives this phenomenon would also be observable to spectroscopic observations through its effect on redox-sensitive species in the atmosphere (Fig. \ref{fig:3}). These properties are encoded in current exoplanet observations, subtly influencing the structure of the radius valley \citep{Fulton2017AJ, Cloutier2020AJ} as observed with Kepler, TESS, and the upcoming PLATO mission. In particular, reflation could manifest as an excess population of puffy super-Earths relative to predictions from standard thermal evolution and photoevaporation models which neglect compositional changes \citep[e.g.][]{Tang2024ApJ, Cherubim2024ApJ}. 

Obtaining comprehensive characterisations on outlier objects that cannot be explained by standard evolution models (i.e., their age, radius, mass, and atmospheric composition) would provide avenues for verifying whether reflation occurs or not. The most promising targets for detecting reflation are super-Earth mass planets with relatively low densities, and moderate to high irradiation levels; particularly those orbiting stars with well-determined ages. Such planets, near the radius valley, and around older systems are of particular interest, as reflation is predicted to occur after significant atmospheric evolution has already taken place.

Interpreting individual cases remains challenging due to degeneracies between formation history, atmospheric escape, and internal processes.  Ultimately, the more powerful approach to use transient probes of planetary evolution is to search for population-level trends. The expected synergies from the upcoming PLATO \citep{Rauer2025ExA} and Ariel \citep{Tinetti2022EPSC} missions will play a key role, providing a statistically significant sample of atmospheres with homogeneous atmospheric constraints. By correlating atmospheric composition (and inferred mean molecular weight) with bulk properties such as radius, density, irradiation, and stellar age, it might be possible to identify systematic deviations from standard evolutionary models. In particular, population-level trends linking atmospheric composition to position relative to the radius valley would provide strong indirect evidence for reflation. Such statistical approaches can also help overcome the limitations of individual-object studies and offer a more robust pathway to disentangling the many relevant evolutionary processes.

\section{Conclusions}

Using coupled interior--atmosphere evolution simulations we demonstrate that the feedback between hydrodynamic escape and redox-controlled outgassing can produce transient re-inflation events of evolved super-Earth atmospheres, a mechanism we term \emph{reflation} (illustrative sketch in Figure \ref{fig:1}). We consider irradiated rocky planet scenarios spanning diverse physical regimes of mantle redox state, initial volatile inventory, escape efficiency, and orbital distance. From these, we identify the conditions under which non-monotonic density evolutionary behaviours arise, and characterise their potentially observable fingerprint. Our conclusions are as follows:
\begin{figure}[t!]
\centering
\includegraphics[width=0.45\textwidth]{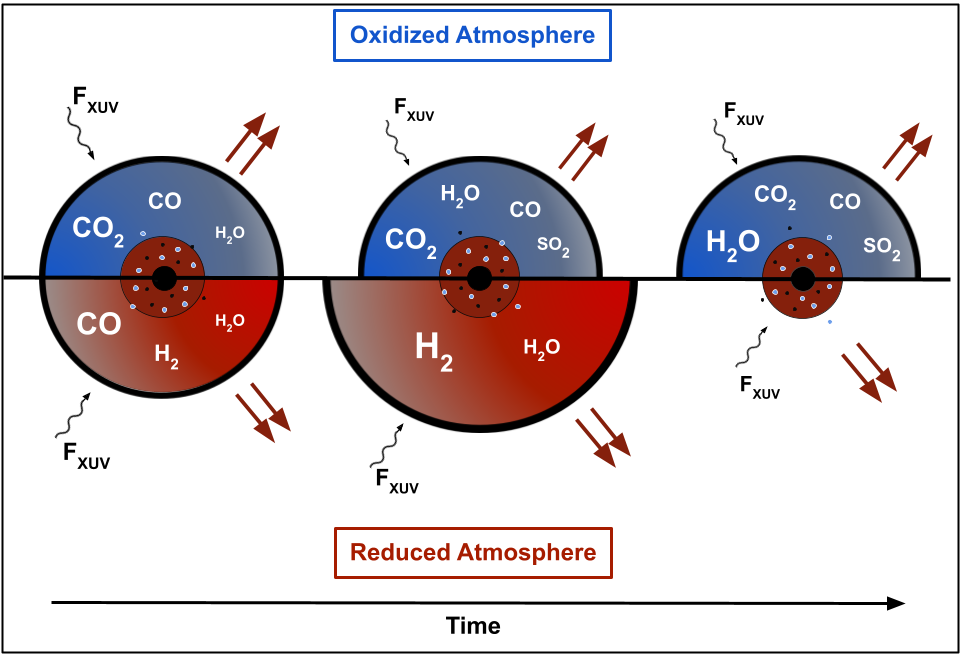}
\caption{Qualitative illustration of the \emph{reflation} mechanism, comparing the standard atmospheric evolution observed in oxidised atmospheres \emph{(top)} and reflation events \emph{(bottom)} observed in reduced atmospheres. Oxidised super-Earths would experience monotonic deflation as their secondary atmospheres are heavy, while reduced interiors favour transient reflation events, arising from the transition from a carbon-rich CO-dominated atmosphere, to an H$_2$-dominated atmosphere by outgassing from the interior.}
\label{fig:1}
\end{figure}
\begin{enumerate}
    \item Geochemically-reduced mantles (near the iron--w\"ustite buffer) initially outgas CO-dominated atmospheres while hydrogen-bearing species are initially dissolved. However, hydrodynamic escape strips outgassed species, so the atmosphere becomes increasingly hydrogen-dominated and lowers the atmospheric MMW, inflating the atmospheric scale height, and transiently reducing bulk planet densities by up to $\sim$ 60\%.
    \item Oxidized Earth-similar mantles do not generate reflation events. Their atmospheres remain dominated by heavy species (mainly CO$_2$ and SO$_2$), so escape leads to monotonic deflation without a significant compositional shift. Such compact atmospheres can be resilient toward total erosion and may survive over Gyrs.
    \item Reflation is favored when escape acts with intermediate efficiencies, at high irradiations, and for initial hydrogen inventories exceeding $\sim$ 5 Earth oceans. Negligible escape will fail to remove the initial heavy atmosphere, while substantial escape strips volatiles entirely before the compositional transition can inflate the atmosphere.
    \item Reflation is confined to the super-Earth regime: planets whose observed compositions are secondary, rather than of nebular H$_2$/He in origin. Reflation offers a direct observational connection between atmospheric properties and interior geochemistry, breaking the degeneracy between composition and thermal state in bulk density measurements.
    \item Population-level surveys correlating atmospheric composition, bulk density, irradiation, and stellar age (e.g., as synergy between PLATO and Ariel) represent the most promising pathway to detecting reflated super-Earths -- thereby constraining deep mantle redox states across the exoplanet population.
\end{enumerate}

The reflation mechanism suggests that atmospheric evolution on rocky exoplanets need not be a one-way street toward stripping. The geochemical and formation history of a planet's interior can, under these identifiable conditions, be written into its atmosphere: an imprint that the next generation of observatories is poised to read.

\section*{Acknowledgments}
We offer our thanks to the anonymous referee, whose review helped to strengthen our article. We are grateful to Emma Postolec for insightful discussions about atmospheric escape. This research was supported by the Branco Weiss Foundation, the European Research Council (ERC) under the European Union's Horizon Europe research and innovation programme (MagmaWorlds, 101219807), the Alfred P. Sloan Foundation (AEThER, G-2025-25284), NASA’s Nexus for Exoplanet System Science research coordination network (Alien Earths, 80NSSC21K0593), and the NWO NWA-ORC PRELIFE Consortium (NWA.1630.23.013). MA is supported by the Swiss National Science Foundation through the Postdoc.Mobility fellowship, grant number 230229. HN acknowledges support from STFC grant UKRI1184. We thank the Center for Information Technology of the University of Groningen for their support and for providing access to the H\'abr\'ok high performance computing cluster. 

\bibliography{references}{}
\bibliographystyle{aasjournalv7}

\end{document}